\documentclass[
    aps,
    prl,                
    reprint,            
    superscriptaddress,
    amsmath,
    amssymb,
    longbibliography,
    nofootinbib
]{revtex4-2}

\usepackage[T1]{fontenc}
\usepackage[english]{babel}
\usepackage{graphicx}
\usepackage{mathtools}
\usepackage{bm}
\usepackage{dcolumn}
\usepackage{float}
\usepackage{tabularx}
\usepackage{hyperref}

\usepackage{comment}

\newcommand*\mean[1]{\bar{#1}} 

\begin{document}

\title{Extended Single-Atom Tweezer Arrays in High-Cooperativity Cavity-QED}

\author{Thomas Picot}
\affiliation{Laboratoire Kastler Brossel, ENS-Université PSL, Sorbonne Université, Collège de France, 24 rue Lhomond, 75005 Paris, France}

\author{Clément Raphin}
\affiliation{Laboratoire Kastler Brossel, ENS-Université PSL, Sorbonne Université, Collège de France, 24 rue Lhomond, 75005 Paris, France}

\author{Marcel Kern}
\affiliation{Laboratoire Kastler Brossel, ENS-Université PSL, Sorbonne Université, Collège de France, 24 rue Lhomond, 75005 Paris, France}

\author{Pierre-Antoine Bourdel}
\altaffiliation[Present address: ]{Pasqal SAS. 24, rue Émile Baudot, 91120 Palaiseau, France}
\affiliation{Laboratoire Kastler Brossel, ENS-Université PSL, Sorbonne Université, Collège de France, 24 rue Lhomond, 75005 Paris, France}

\author{Théo Ahamdach}
\affiliation{Laboratoire Kastler Brossel, ENS-Université PSL, Sorbonne Université, Collège de France, 24 rue Lhomond, 75005 Paris, France}

\author{Jakob Reichel}
\affiliation{Laboratoire Kastler Brossel, ENS-Université PSL, Sorbonne Université, Collège de France, 24 rue Lhomond, 75005 Paris, France}

\author{Romain Long}
\email[Corresponding author: ]{long@lkb.ens.fr}
\affiliation{Laboratoire Kastler Brossel, ENS-Université PSL, Sorbonne Université, Collège de France, 24 rue Lhomond, 75005 Paris, France}

\date{\today}


\begin{abstract}
   A central challenge for cavity-QED-based quantum technologies is to make high-cooperativity optical interfaces compatible with site-resolved arrays of single atoms. Here, we demonstrate optical tweezer arrays of individual $^{87}$Rb atoms inside a fiber Fabry-Perot microcavity with single-atom cooperativity $\mathcal{C} \sim 90$. We combine background-free site-resolved fluorescence imaging of extended arrays with collective coupling to a common cavity mode for arrays with a mean atom number up to $\mean{N}\simeq36$. These results establish a high-cooperativity platform for many-body cavity-QED with site-resolved detection and control.
\end{abstract}

\maketitle



Over the last decades, cavity quantum electrodynamics (cavity-QED)  has evolved from a seminal framework for exploring quantum light-matter interaction and entanglement \cite{ Kimble1998,Haroche2006} into a versatile platform for quantum technologies. 
Combined with cold atoms, optical cavity-QED now plays a central role in a broad range of applications, including quantum networks and memories \cite{Reiserer2015, Brekenfeld2020}, spin-squeezed states for quantum metrology \cite{Leroux2010, Hosten2016}, long-range spin models for quantum simulation \cite{Davis2019, Defenu2023, Sauerwein2023, Kroeze2025} and cavity-mediated gates or interconnects between quantum processors \cite{Welte2018, Ramette2022c, Covey2023a, Sinclair2025}. 
Until recently, two main experimental approaches have been explored: single emitters strongly coupled to a cavity and large ensembles of emitters with weak single-particle coupling but strong collective coupling to a resonator.

Following pioneering two-atom experiments \cite{Reimann2015a, neuznerInterferenceDynamicsLight2016}, a new generation of cavity-QED platforms is now emerging, combining optical cavities with tweezer arrays of single atoms \cite{ Deist2022c, Deist2022, Liu2023, Hartung2024b, Seubert2025, Grinkemeyer2025, Wang2025c, Hu2025, Shaw2026, Santis2026}. This approach combines local control and detection of individual atoms with the collective coupling of many emitters to a common cavity mode. 
Until very recently, however, this combination had remained mostly limited to a moderate-cooperativity regime: demonstrations reaching cooperativities on the order of ten or larger have involved only up to two individually controlled atoms \cite{Grinkemeyer2025}. Extending high-cooperativity cavity-QED to larger, spatially resolved arrays is therefore an important step toward many-body cavity-QED with microscopic control.

Here, we demonstrate optical tweezer arrays of single atoms strongly coupled to a high-finesse fiber Fabry–Perot microcavity, combining strong single-atom coupling with site-resolved detection. We first characterize the atom-cavity interface by mapping the cavity mode with a single atom, maximizing the coupling and extracting a single-atom cooperativity $\mathcal{C} \simeq 88(4)$.
We then operate the system in the single-atom regime to demonstrate cavity-based hyperfine-state detection \cite{Gehr2010} with a fidelity of $99.4(3)\%$. By repeating this characterization with a single tweezer positioned at locations corresponding to the extremal sites of a  $20\times5$-site tweezer array, we obtain fidelities above $98.8(5)\%$, showing that cavity-based detection is compatible with site-resolved readout over the full spatial extent of the array.

While cavity-based qubit detection enables non-destructive readout \cite{Volz2011}, entanglement generation \cite{Haas2014}, and mid-circuit measurements \cite{Deist2022, Hu2025}, it is inherently sequential and therefore becomes time-consuming for large qubit arrays. When non-destructive detection is not required, fluorescence imaging offers a naturally parallelized approach to multi-qubit detection with direct access to the spatial configuration of the array.
In optical microcavities, however, strong coupling comes at the cost of placing atoms close to macroscopic surfaces, less than $70\,\mu$m from the fiber mirrors in our experiment. Scattering of the fluorescence beams from these surfaces and nearby mounts produces a large fluctuating background that can hinder single-atom detection.
To overcome this limitation, we implement a background-free fluorescence scheme for $^{87}\mathrm{Rb}$ atoms based on two-photon excitation  \cite{Mcgilligan2020a, Menon2024} and characterize both the detection fidelity and the survival probability during imaging. This enables us to generate and image atom arrays tailored to the cavity mode profile, in configurations designed for homogeneous atom-cavity coupling or for controlled coupling inhomogeneity.

Finally, we couple the arrays to the cavity mode and observe collective enhancement of the atom-cavity coupling for increasingly extended geometries: from one-dimensional chains optimized for maximum coupling to  homogeneously coupled two-row array and to $20\times4$-site configurations. This realizes multi-row tweezer arrays in which high single-atom cooperativity is combined with collective coupling to a common cavity mode. With a mean atom number up to $\mean{N}\simeq36$, these results establish a platform for many-body cavity-QED  with programmable atomic configurations and local readout.

\begin{figure*}[t!]
    \centering
    \includegraphics[width=1\linewidth]{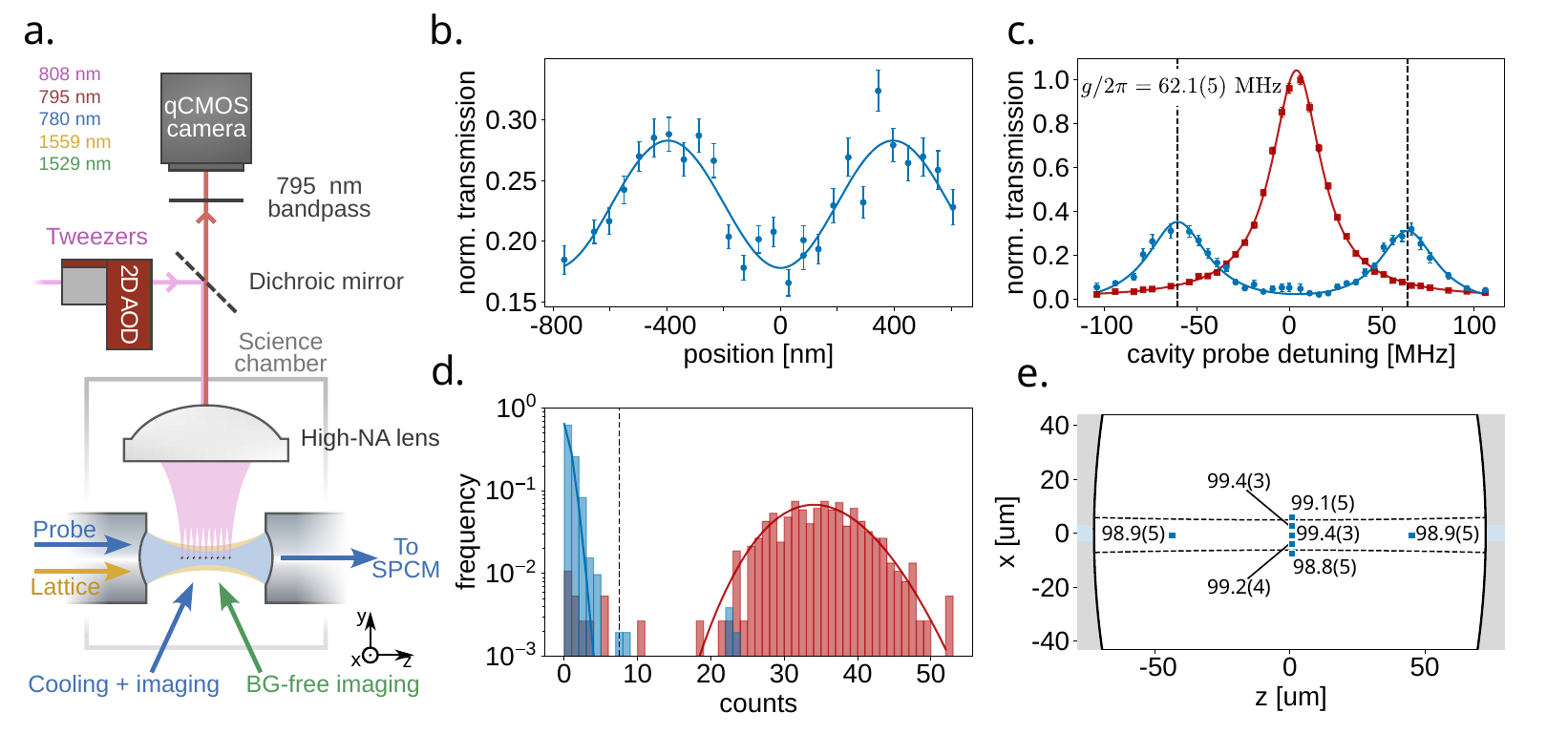}
    \caption{\textbf{Single atom strongly coupled to the cavity}.
        \textbf{a.} Simplified sketch of the experiment. Atoms are trapped inside a fiber Fabry-Perot microcavity by an array of optical tweezers generated by a pair of acousto-optic deflectors (2D AOD) along the cavity axis, and focused with a high N.A. lens.
        \textbf{b.} Mapping of the cavity mode by moving the single-atom tweezer along the $\mathbf{z}$ axis and measuring cavity transmission while measuring the atom position, and sinusoidal fit (solid line). The cavity probe frequency sits on the slope of the polariton.
         \textbf{c.} Vacuum Rabi Splitting observed in cavity transmission for a single atom at the center of the cavity (blue points), fitted by a double Lorentzian (blue curve) yielding $g/2\pi = 62.1(5)$ MHz and $C = g^2/\kappa\gamma=88(4)$. For reference, cavity transmission when no atom is present (red squares) with Lorentzian fit (red curve).
        \textbf{d.} Histogram of the photon counts, collected during 100$\,\mu$s of probing, with Poissonian fits (solid lines) and optimal threshold (dotted line). Blue corresponds to a preparation in F = 1, while red corresponds to a preparation in F = 2.
        \textbf{e.} Mapping of the readout fidelity across the spatial extents of a $20 \times 5$-site array.}
    \label{fig:Fig1}
\end{figure*}

\section*{Single-Atom Strong Coupling and Cavity-Based Qubit Readout}

The core experimental apparatus, based on $^{87}\mathrm{Rb}$ atoms coupled to a fiber Fabry-Perot microcavity, has been described previously \cite{Baghdad2023}. Briefly, the two-wavelength fiber cavity with negligible birefringence \cite{Garcia2018a, Garcia2020} allows us to reach the high-cooperativity regime, with a maximum theoretical single-atom cooperativity of $\mathcal{C} = g^2/(\kappa \gamma)=132$. Here $g$ is the single-atom single-photon coupling strength, while $\kappa$ and $\gamma$ are the the cavity and atomic HWHM decay rates, respectively, with $(g, \kappa, \gamma)/2\pi = (75.0, 14.2, 3.0)$ MHz.
Cold atoms are first produced in a magneto-optical trap and transported into the cavity region using an optical dipole beam \cite{Ferri2022}.  They are then transferred into an optical tweezer array by ramping down the transport beam, while ramping up the $808\,$nm tweezer light (see Fig. \ref{fig:Fig1}a). During loading and cooling, the tweezer array is located $20\,\mu$m away from the cavity axis, in order to avoid any potential detrimental effects associated with the cavity.
The atoms are then moved into the cavity within $1.5\,$ms, where an intracavity lattice at 1559$\,$nm, commensurate with the probe field, is ramped up to form a hybrid trap with the tweezers \cite{Garcia2020}. This lattice pins the atoms at the antinodes of the probe cavity mode, making the atom-cavity coupling less sensitive to residual position jitter of the tweezer array and providing stronger confinement along the cavity axis \cite{Bourdel2022, Hartung2024b}.
Throughout this study, the cavity is resonant to the $5S_{1/2}, |F = 2 \rangle \to  5P_{3/2},|F' = 3 \rangle$ transition.
As the cavity shows no measurable birefringence, we probe the coupled system with $\sigma^+$-polarized light, realizing an effective two-level vacuum Rabi splitting that is not affected by the excited-state hyperfine structure \cite{Gehr2010}. In the single-atom regime, this also efficiently pumps the atom into the stretched Zeeman sublevel $5S_{1/2},|F=2, m_F=2\rangle$.

\begin{figure*}[t!]
    \centering
    \includegraphics[width=1\linewidth]{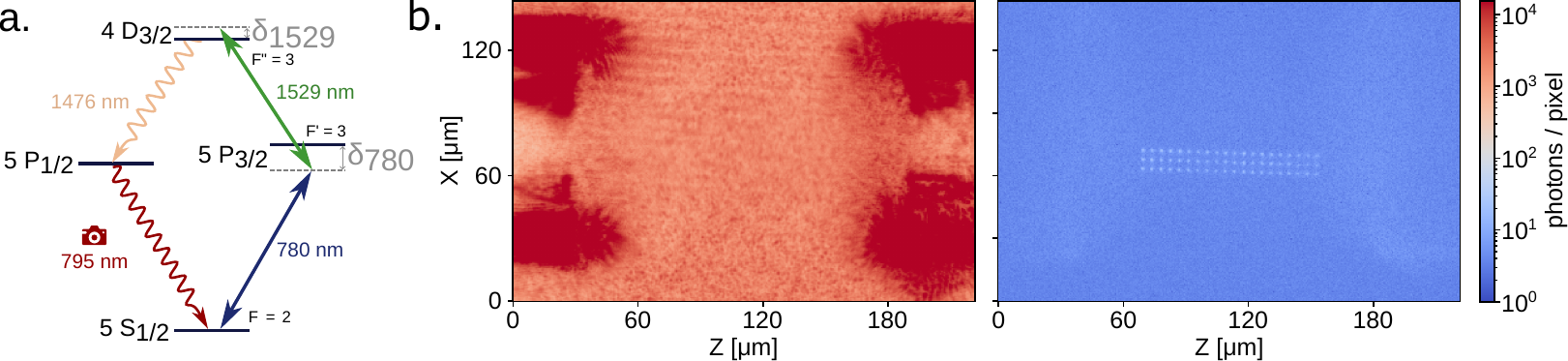}
    \caption{\textbf{Background-free fluorescence scheme}.
        \textbf{a.} Relevant level scheme. The atoms are excited via the D2 transition at 780 nm (blue), followed by a second excitation at 1529 nm (green). The atom subsequently decays to an intermediate state at 1476 nm (yellow), before emitting a photon on the 795$\,$ nm imaging transition.
        \textbf{b.} Comparison between standard fluorescence imaging and the background-free scheme with 50 ms exposure time. Left: Standard single-photon excitation at low power ($s_{780}=0.1$ per beam) to avoid camera saturation. Right: Background-free fluorescence image obtained with $s_{780}=3.5$ and $s_{1529}=5.2$ per beam. 
        For standard fluorescence imaging, the residual background in a $3\times3$-pixel ROI sums up to $20000$ photons and has shot-to-shot fluctuations of $1700$ photons, larger than the expected single-atom signal. For the background-free scheme, these values are reduced to 13 and 6 photons, respectively.
    }
    \label{fig:Fig2}
\end{figure*}

We benchmark the atom-cavity system in the single-atom regime by loading a single optical tweezer and performing a standard vacuum Rabi splitting measurement. 
To avoid the large light shifts induced by the intracavity lattice at 1559$\,$nm \cite{Baghdad2023}, we chop the lattice and probe light out of phase. 
To maximize the atom-cavity coupling, we map the cavity mode by moving the optical tweezer along the three spatial axes (see Fig.~\ref{fig:Fig1}a), while probing on the slope of a polaritonic resonance. Translating the tweezer shifts the center of the hybrid trap, allowing us to map the atom-cavity coupling and, in particular, to resolve the standing wave of the probe field (see Fig.~\ref{fig:Fig1}b). 
From the vacuum Rabi splitting shown in Fig.~\ref{fig:Fig1}c, we extract $g/2\pi=62.1(5)\,$MHz, corresponding to a single-atom cooperativity of $\mathcal{C}= 88(4)$.
We attribute the discrepancy with the theoretical maximum mainly to thermal motion, including possible heating during the ramp-up of the intracavity lattice.

Beyond vacuum Rabi splitting spectroscopy, we leverage the high atom-cavity cooperativity as a state-selective readout channel for cavity-assisted hyperfine-state detection, distinguishing atoms in  $|F=2\rangle$ from uncoupled atoms in $|F=1\rangle$. Figure~\ref{fig:Fig1}d shows the measured qubit-state cavity transmission histograms for a detection time of 100$\,\mu$s, fitted with a double Poissonian distribution. The detection fidelity is obtained by setting an optimal threshold between the two distributions and evaluating the corresponding false-positive and false-negative probabilities. 
For atoms prepared in  $|F=2\rangle$, we obtain a state-detection fidelity of $99.4(3)\%$, mainly limited by the $60\,$ms hyperfine-state effective lifetime.
For atoms prepared in $|F=1\rangle$, the fidelity is $97.3(8)\%$, constrained by the efficiency of the blast pulse used to remove atoms remaining in $|F=2\rangle$ before the detection. This preparation imperfection is visible as residual counts for atoms in $|F=1\rangle$ (blue) overlapping with the $|F=2\rangle$ distribution (red). The two Poissonian distributions are otherwise well resolved, with an inferred overlap negligible compared to the measured state-preparation and loss errors. Hence, the observed fidelity is limited by state preparation and survival rate rather than by intrinsic readout discrimination. 

We repeat the same qubit-state measurement at different tweezer positions in order to probe the spatial extent of a $20 \times 5$-site target array.  Along the transverse direction, the tweezer is placed at the positions corresponding to the five rows of the array, spaced by $3.2\,\mu$m. This array geometry fits within the cavity mode and yields theoretical single-atom cooperativities of approximately $131$, $71$ and $11$ for the central, intermediate and outer lines, respectively (see Fig.~\ref{fig:Fig1}e). Even at the outermost transverse positions, we find a detection fidelity for atoms in $|F=2\rangle$ exceeding $98.8\%$.
We also test the longitudinal homogeneity of the  $|F=2\rangle$-state detection by placing the tweezer at the two extremal positions of the target array along the cavity axis. These positions are more than $40\,\mu$m away from the cavity center and are chosen to be displaced by an integer number of periods of the intracavity lattice, so that they correspond to antinodes of both the probe and trapping standing waves. At these positions, we measure a $|F=2\rangle$ state detection fidelity of $98.9(5)\%$.
Although only one tweezer is present during this characterization, the probed positions correspond to the transverse and longitudinal edges of the target array.  Together with selective site readout \cite{Hu2025} and standard atom rearrangement techniques, this demonstrates that cavity-assisted state detection remains compatible with the spatial extent of hundred-site tweezer geometries.

\section*{Background-free fluorescence imaging of single-atom arrays in a fiber microcavity}

\begin{figure*}[t!]
    \centering
    \includegraphics[width=1\linewidth]{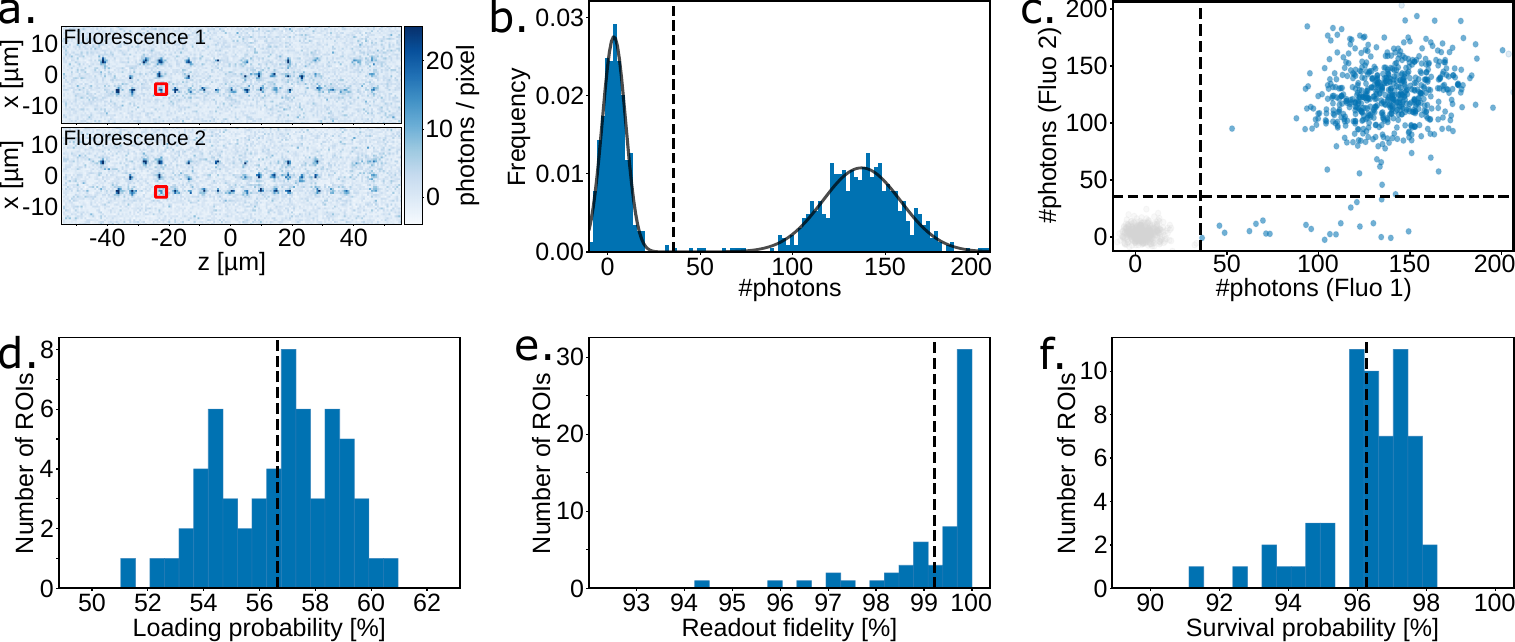}
    \caption{\textbf{Fluorescence characterization}.
        \textbf{a.} Two successive single-shot fluorescence images of a $20 \times 3$ tweezer array taken with 50 ms exposure time. The upper image and the lower image are separated by a 25 ms cooling phase. Red rectangle indicates a 3$\times$3 pixel  ROI.
        \textbf{b.} A histogram of the summed photon counts in the ROI over 1000 experimental cycles, with a double gaussian fit (solid line) and computed optimal threshold (dashed line).
        \textbf{c.} Scatter plot of the summed photon counts in the two fluorescence images with optimal threshold (dashed line). This figure shows the survival probability of an atom after the first fluorescence.
        \textbf{d.} Histogram of the loading probability over the 60 tweezer sites. The averaged value (dashed line) gives $56(2)\%$ of loading probability.
        \textbf{e.} Histogram of the readout fidelity over the 60 tweezer sites, exhibiting a skewed distribution with an average value (dashed line) of $99(1)\%$.
        \textbf{f.} Histogram of the survival probability over the 60 tweezer sites, with an average value (dashed line) of $96(1)\%$.} 
    \label{fig:Fig3}
\end{figure*}

When non-destructive cavity-based detection is not required, for instance during initialization or final readout, fluorescence imaging offers a  parallel approach to multi-qubit detection and is the method of choice in many experiments with single-atom arrays and quantum gas microscopes. Yet, for cold atoms trapped inside optical microcavities, or in the vicinity of other photonic interfaces, scattering of the fluorescence beams from nearby surfaces can produce a large fluctuating background, preventing atom-presence detection across the array.
To circumvent this limitation, we implement a background-free fluorescence scheme relying on the multilevel structure of alkali atoms \cite{Mcgilligan2020a, Menon2024}.  The $^{87}$Rb atoms  are excited by two photons at $780\,$nm and $1529\,$nm, promoting them from the  $5S_{1/2}, F =2$ ground level to the $4D_{3/2}, F'' =3$ state, as shown in Fig.~\ref{fig:Fig2}a. Then, they can decay through the $5P_{3/2}$ state, but also through the $5P_{1/2}$ state with a favorable branching ratio of $5/6$ \cite{Noh2012}. Therefore, this diamond scheme allows us to image the atoms on the $D_1$ line, while efficiently filtering out excitation light. In addition, 
the detected wavelength at $795\,$nm remains close to the design wavelength of the imaging lens and to the tweezer wavelength of 808$\,$nm, avoiding large chromatic shifts.
Fig.~\ref{fig:Fig1}a shows a simplified sketch of the fluorescence setup. Atoms are trapped in an array of optical tweezers tightly focused near the cavity mode by a high-numerical-aperture lens (NA=0.5) \cite{Ferri2020}. The same lens collects the fluorescence light and enables single-atom-resolved imaging on a qCMOS camera.

To assess the suppression of the background noise, we first compare the background-free scheme with standard fluorescence imaging in Fig.~\ref{fig:Fig2}b. The first image is obtained with single-photon excitation at 780$\,$nm and low optical power. Even in this configuration, the fiber regions are close to saturation. At the powers required for efficient fluorescence imaging, the shot-to-shot fluctuations of the residual background in the central region of the cavity exceed the single-atom signal expected in a $3 \times 3$-pixel region of interest (ROI), corresponding to a single tweezer site, as shown in Fig.~\ref{fig:Fig2}b. 
In contrast, the background-free scheme strongly suppresses this fluctuating background, allowing us to image a $20\times3$ tweezer array with a spacing of $4.68\,\mu$m (Fig.~\ref{fig:Fig2}b).

\begin{figure*}[t!]
    \centering
    \includegraphics[width=\linewidth]{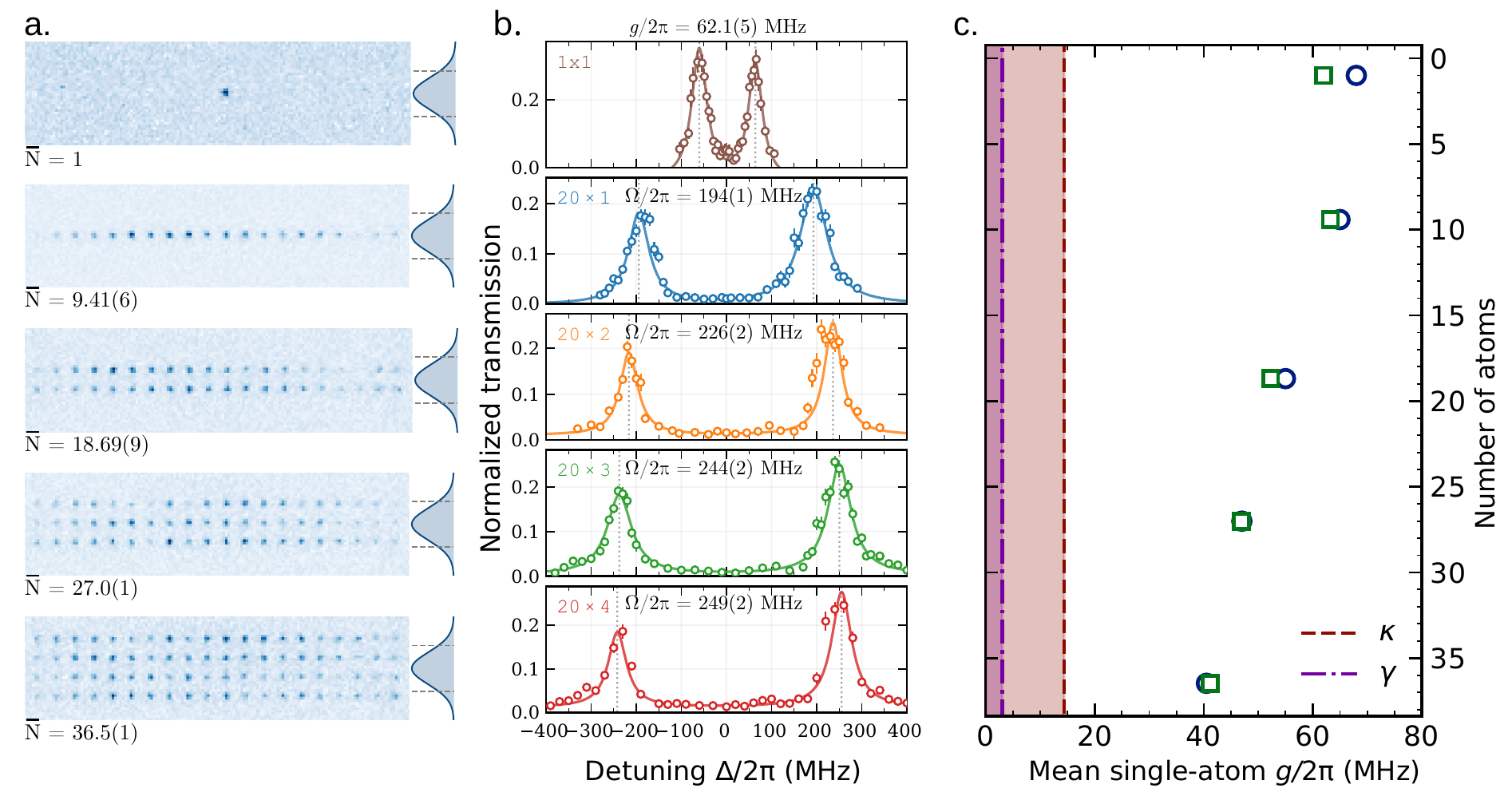}
    \caption{
        \textbf{Vacuum Rabi-splitting measurement for different configurations of single-atom array.}
        \textbf{a.} Averaged fluorescence image of the tweezers configurations. Gaussian shape on the right indicates spatial extension of the cavity mode of waist $w_0 = 5.6 \;\textrm{µm}$. The indicated number $\mean{N}$ corresponds to the averaged number of atoms during the probing sequence.
        \textbf{b.} Vacuum Rabi Splitting for the different array configurations. Each point is an average of around 30 experimental cycles. Separation between the two polaritonic peaks $2\Omega$ is evaluated by a double Lorentzian fit (solid line).
        \textbf{c.} Comparison between the expected average single-atom coupling strength (blue circles) with the measured value $g = \Omega/\sqrt{\mean{N}}$ (green squares).
    }
    \label{fig:fig4}
\end{figure*}

Having established the strong suppression of background fluctuations, we now quantify the detection fidelity and atom survival rate during fluorescence imaging. Figure~\ref{fig:Fig3}a shows representative single-shot images of the tweezer array with a stochastic loading probability of $\sim 56\%$. By accumulating more than one thousand realizations, we obtain the histogram shown in Fig.~\ref{fig:Fig3}b, corresponding to the photon counts collected in a $3\times3$-pixel ROI for a selected site. The histogram shows two well-separated peaks, representing an empty and occupied tweezer site. From the separation between the two distributions, we determine an optimal threshold and obtain a detection fidelity of $99.9(1)\%$ for this site.   

To determine the survival probability, we perform two successive fluorescence measurements. Figure~\ref{fig:Fig3}c compares the number of photons detected during the first and second fluorescence images. Using the optimal threshold, four different regions are defined. The lower-left region corresponds to empty tweezer sites, the upper-right region as atoms detected in both images, and  the lower-right region as atoms detected only during the first image but lost before the second one. For the selected site, we obtain a survival probability of $96.3(3)\%$. This value is mainly limited by the finite lifetime of the atoms in the tweezers. In contrast, events in the upper-left region, corresponding to atoms detected only in the second image, are rare.

Figures \ref{fig:Fig3}d-f show the loading probability, readout fidelity and survival probability for each site of the array. Averaged over the array, we find a loading probability of $56(2)\%$, a detection fidelity of $99(1)\%$, and a survival probability of $96(1)\%$. The tweezer intensities are independently measured on a CMOS camera, placed on a separate diagnostic path, yielding an initial site-to-site inhomogeneity of $\sim 14\%$, reduced to $\sim 1\%$ after optimization. The residual variations observed across the tweezer array are likely not limited by the tweezer intensity profile itself, but rather by spatial inhomogeneities of the cooling light, for instance due to interference between the molasses beams.  

By jointly optimizing the detection fidelity and the survival probability, we determine the optimal excitation parameters used for fluorescence imaging, reaching a detected photon rate of 1-2$\,$kHz. This moderate rate reflects a trade-off: efficient cooling favors larger detunings, whereas a high scattering rate requires excitation close to resonance. Compared with  standard single-photon fluorescence imaging, this compromise might be harder to optimize here, given that cooling and fluorescence generation involve different branches of the diamond excitation scheme.

Still, the strong suppression of the background fluctuations compensates for the moderate photon rate, allowing us to achieve high detection fidelity, while preserving a large survival probability for a typical exposure time of $50\,$ms. Overall, the scheme provides a robust, low-loss, and high-fidelity fluorescence readout for atoms trapped in tweezer arrays close to optical surfaces. Therefore, it offers an important tool for combining single-atom-resolved imaging with strong coupling to optical microcavities or other photonic structures.


\section*{High-Cooperativity Cavity-QED with Single-Atom Arrays}

Having demonstrated single-atom strong coupling in the fiber Fabry-Perot microcavity and site-resolved fluorescence imaging of extended tweezer arrays, we now combine these two capabilities and couple different array geometries to the mode of the cavity. As the number of atoms can reach several tens, we include an optical pumping phase to prepare the atoms in the stretched state $|F=2, m_F=2\rangle$, before probing the cavity transmission.

We first consider a 20-site tweezer chain spaced by $4.68\,\mu$m and positioned at the center of the transverse cavity mode (Fig.~\ref{fig:fig4}a). For a mean atom number of $\mean{N}=9.41(6)$, we measure the collective vacuum Rabi splitting presented in Fig.~\ref{fig:fig4}b. A fit to the spectrum yields a mean single-atom cooperativity $\overline{\mathcal{C}}=92(3)$, close to the value measured with a single atom. This confirms that the atoms in the chain are trapped close to the antinodes of the standing wave formed by the intracavity probing light at $780\,$nm and thus near-optimal coupling along the chain.

We then extend the measurement to multi-row  single-atom arrays, with the same tweezer spacing of $4.68\,\mu$m along both directions. In a $20\times2$ configuration, with the long direction along the cavity axis, the two rows are placed symmetrically around the transverse Gaussian cavity mode center. This geometry provides a nearly homogeneous coupling of all atoms to the cavity mode. From the measured spectrum and with $\mean{N}=18.69(9)$ atoms, we extract a mean single-atom cooperativity of $\overline{\mathcal{C}}=63(2)$. The reduction in cooperativity is consistent with the expected decrease of the coupling strength away from the center of the Gaussian cavity mode.

Finally, we investigate increasingly extended multi-row arrays with controlled spatial inhomogeneity of the atom-cavity coupling. We realize $20 \times 3$ and $20 \times 4$-site configurations, containing on average $\mean{N}=27.0(1), 36.5(1)$ atoms, respectively. For the largest atom number, we extract a mean single-atom cooperativity $\overline{\mathcal{C}}=39(1)$. Although the coupling varies substantially between the central row and the outermost rows, the cooperativity remains above 5 at every site. The observed reduction of the average coupling strength is consistent with the Gaussian transverse profile of the cavity mode as shown in Fig.~\ref{fig:fig4}c.
The observed asymmetry between the two polariton peaks may arise from imperfect preparation in the stretched state and/or from a small impurity of the probe polarization. In this case, atoms can weakly couple to other hyperfine levels of the excited state, in particular $F'=2$. This distorts the ideal two-level vacuum Rabi spectrum and leads to asymmetric peak amplitudes. 

The number of coupled atoms could be further increased by performing a rearrangement procedure before transferring the atoms into the cavity mode \cite{Endres2016a,Barredo2016}, by reducing the tweezer spacing and by extending the array length within the cavity mode volume. These improvements should enable larger atom numbers, while preserving high single-atom cooperativity and site-resolved detection across the array. A straightforward extrapolation based on the measured cavity-mode profile indicates that $\sim60$ atoms could be coupled nearly homogeneously with $\overline{\mathcal{C}}>60$, while configurations containing up to $\sim150$ atoms appear realistic in a controlled inhomogeneous-coupling regime.

\vspace{0.5cm}


In conclusion, we have combined single-atom strong coupling in a fiber Fabry-Perot microcavity with high-fidelity, low-loss, background-free fluorescence imaging of extended single-atom arrays. This brings extended chains and multi-row arrays of single atoms into the high-cooperativity regime, enabling the collective coupling of tens of atoms to a common cavity mode. 
Together with programmable atomic configurations and site-resolved imaging, this opens a route toward many-body cavity-QED with microscopic control. This approach is directly relevant to quantum computing and networking, including cavity-mediated gates, mid-circuit measurements, and interconnects between neutral-atom processors \cite{Jandura2024a,Deist2022,Hu2025,Ramette2022c,Covey2023a}. In parallel, it offers perspectives for multiparameter quantum metrology \cite{Pezze2018a,Gessner2020, Baamara2023,  Li2026} and for the quantum simulation of transport and localization phenomena in long-range disordered spin systems \cite{Botzung2020a, Chavez2021b}.

\emph{Note added:} While finalizing this manuscript, we became aware of independent related work reporting optical tweezer arrays coupled to a high-cooperativity Fabry--Perot cavity \cite{Ye2026}.
\vspace{0.5cm}


\begin{acknowledgments}
\textbf{Acknowledgments}
This project has received funding from: Agence Nationale de la Recherche (CLIMAQS project, ANR-14-CE32-0002) and in the framework of France 2030 (QUTISYM project, ANR-23-PETQ-0002).
\end{acknowledgments}


\bibliography{2026_Cavity_Atom_Array}

\end{document}